\documentclass[prd,onecolumn,notitlepage,showpacs,preprintnumbers,amsmath,amssymb,nofootinbib,APS,10pt,superscriptaddress]{revtex4-1}

\usepackage{dcolumn}
\usepackage{bm}
\usepackage[applemac]{inputenc}
\usepackage[spanish,english]{babel}
\usepackage{amsmath,amssymb,amsfonts,latexsym,cancel}
\usepackage{graphicx}
\usepackage{color}
\usepackage{soul}
\usepackage{ulem}
\usepackage{hyperref}

\allowdisplaybreaks

\newcommand{\be}{\begin{equation}}
\newcommand{\ee}{\end{equation}}
\newcommand{\bea}{\begin{eqnarray}}
\newcommand{\eea}{\end{eqnarray}}

\begin{document}
\title{{\bf 
Pair creation in electric fields, anomalies, and renormalization of the electric current}}

\author{Antonio Ferreiro}\email{antonio.ferreiro@ific.uv.es}
\author{Jose Navarro-Salas}\email{jnavarro@ific.uv.es}
\affiliation{Departamento de Fisica Teorica and IFIC, Centro Mixto Universidad de Valencia-CSIC. Facultad de Fisica, Universidad de Valencia, Burjassot-46100, Valencia, Spain.}

\begin{abstract}

We investigate the Schwinger pair production  phenomena in spatially homogeneous strong electric fields.  We first consider scalar QED in four-dimensions and discuss the potential ambiguity in the adiabatic order assignment  for the electromagnetic potential required to fix the renormalization subtractions. We argue that this ambiguity can be solved by invoking the conformal anomaly when both electric and gravitational backgrounds are present.  We also extend the adiabatic regularization method for spinor QED in two-dimensions and find consistency with the chiral anomaly. We focus on the issue of the renormalization of the electric current $\langle j^\mu \rangle$ generated by the created pairs. We illustrate how to implement the renormalization of the electric current for the Sauter pulse. \\


{\it Keywords:} Schwinger effect, particle creation, quantum field  theory in external conditions, renormalization, adiabatic regularization. \end{abstract}

\pacs{04.62.+, 11.10.Gh, 12.20.-m, 11.30.Rd} 

\date{\today}
\maketitle

\section{Introduction}\label{Introduction}

A time-dependent gravitational field yields the creation operators of  quantum fields to evolve into a superposition of creation and annihilation operators. This produces the spontaneous creation of particle-antiparticle pairs \cite{parker-toms, Waldbook, fulling, birrell-davies}. This effect was first discovered and widely analyzed 
in the physical context of an expanding universe \cite{parker66, parker68}. 
In the early seventies, a similar transformation between creation and annihilation operators was proved to occur  for accelerated observers in the Rindler wedge of Minkowski space \cite{fulling73} and also  in the spacetime describing a gravitational collapse forming a black hole \cite{hawking74}. 
Subsequent investigations concluded that gravitons are also created  by the expansion of the universe  \cite{ford-parker, Grishchuk}. Shortly after the proposal of the inflationary universe \cite{guth}, the creation of scalar perturbations was analyzed in \cite{inflation2}.  A similar superposition of creation and annihilation operators takes place if the quantized field is coupled to a time-varying scalar  field background. Therefore,  particle creation can be enhanced  after the end of inflation,  when the inflation, regarded as an external scalar field coupled to quantized matter fields, starts to rapidly  oscillate  in time \cite{KLS}.\\ 

In this paper we want to focus on the particle creation phenomena caused by a time varying background gauge field. These electromagnetic processes are strongly motivated by the upcoming high intensity and ultrashort laser experiments \cite{ELI, Dunne0}, which will allow us to understand nonperturbative regimes in QED where vacuum particle creation becomes relevant \cite{Pittrich-Gies}. The most paradigmatic example of this is the pair-production by a strong and spatially homogeneous electric field $\vec E (t) = E(t)\hat x$, where one assumes  a configuration with an initial  vanishing electric field $E(t) \to 0$ as $t \to -\infty$, an intermediate period with a smoothly varying electric field, and a final decay to a zero electric field $E(t) \to 0$ when $t \to +\infty$. A prototype  for this is  the pulsed configuration $\vec E = (E_0 \cosh^{-2}(\omega_0t), 0, 0)$. In the limit $\omega_0 \to 0$ one recovers the constant electric field, which is actually an inherent   assumption in the well-known Schwinger effect \cite{Schwinger51}. Schwinger's derivation, extending previous work by Sauter \cite{Sauter} and Heisenberg and Euler \cite{Heisenberg-Euler}, is obtained within the effective action formalism in quantum electrodynamics (QED). It has been of great interest in theoretical research over many years \cite{Brezin-Itzykson, Popov, Dunne2, Dunne3, Dunneprl08, Piazza, Dunneprl, Adorno, Adorno2, Adorno3} and it  may be at the border of being experimentally verified.\\

A fundamental problem in gravitational  processes is the calculation of the expectation values $\langle T_{\mu\nu} \rangle$, which, in addition to provide definite quantities for energy density, pressure, or radiation fluxes,  would act as the proper source for the  (semiclassical) Einstein's equations. The computations are involved, as we have to deal with ultraviolet (UV) divergences not present in Minkowski space. The corresponding subtractions needed for renormalization requires more sophisticated  methods. 
 Equivalently, in a time-dependent electric field the proper source of the Maxwell equations is the electric current which again possesses UV divergences. In this paper we analyze the renormalization of the electric current $\langle j_\mu \rangle$ for spatially homogeneous electric fields, like the pulsed configuration mentioned above, by importing and extending renormalization methods originally proposed in cosmological scenarios \cite{parker-fulling, Bunch80, Anderson-Parker, LNT, RNT, Ghosh16, RFNT}. The electric current is probably the most important local observable to be consider in a near-future  detection of the Schwinger effect. Therefore, the renormalization of both  observables $\langle T_{\mu\nu} \rangle$ and  $\langle j_\mu \rangle$ merit a detailed and simultaneous analysis. Moreover, new interests in cosmological scenarios of the Schwinger effect \cite{KA, SSX} demand the need of a regularization method involving both electric and gravitational fields. \\
 
 The goal of this paper is to further extend the adiabatic regularization method for scalar \cite{parker-fulling, Bunch80, Anderson-Parker} and Dirac fields \cite{LNT, RNT, Ghosh16, RFNT} living  in a Friedmann-Lemaitre-Robertson-Walker (FLRW) spacetime by adding the interaction with an external homogeneous electric field. The organization of the paper is as follows. In Sec. II we first consider scalar QED in four-dimensions and analyze the  ambiguity in the adiabatic order assignment for the electromagnetic potential. We fix the ambiguity by invoking agreement with the trace anomaly when both electric and gravitational fields are present. In Sec. III we further extend the adiabatic scheme for Dirac fermions in two-dimensions. We find consistency with the axial anomaly. In Sec. IV we illustrate how to evaluate the renormalized electric current for the Sauter pulse in flat space using the improved renormalization methods introduced in Sec. III. Finally, in Sec. V we summarize our conclusions.

\section{Scalar QED in FLRW spacetime, adiabatic expansion, and renormalization}

Let us consider a quantized charged scalar field  in a FLRW metric of the form $ds^2=dt^2-a^2(t)d\vec{x}^2$. The scalar field obeys the Klein-Gordon field equation (we assume a generic coupling $\xi$ to the scalar curvature)
\be (D_\mu D^\mu + m^2+\xi R)\phi =0 \ , \ee
where
\be D_\mu \phi = (\nabla_\mu + iqA_\mu )\phi . \ee
Assuming that the electric field is spatially homogeneous and the magnetic field is zero, we take the electric field in the direction of the $x$ axis. For our purposes it is very convenient to choose a gauge such that only the $x$-component of the vector potential is nonvanishing: $A_{\mu}= (0, -A(t), 0, 0)$. Therefore, the field strength  is  given by $F_{0i}= (-\dot A (t), 0, 0) $ 
and the  Klein-Gordon equation becomes 

\be \ddot \phi -\frac{\vec \nabla^2}{a^2} \phi+3 \frac{\dot{a}}{a}\dot \phi +\frac{2iqA}{a^2} \partial_x \phi + \frac{q^2 A^2}{a^2}\phi + (m^2+ \xi R)\phi=0 \ . \label{motionscalar2} \ee
The quantized field is expanded in Fourier modes as
\be \phi(x)= \frac{1}{\sqrt{2(2 \pi a)^3}}\int d^3 \vec{k} [A_{\vec{k}}e^{i\vec{k}\vec{x}}h_{\vec k}(t)+B_{\vec{k}}^{\dagger}e^{-i\vec{k}\vec{x}}h^*_{-\vec k}(t) ] \ , \label{phisolution2} \ee
where
 $A_{\vec{k}}^{\dagger}, B_{\vec{k}}^{\dagger}$ and $A_{\vec{k}}, B_{\vec{k}}$ are the usual creation and annihilation operators. The mode functions $h_{\vec k}(t)$ must obey the Wronskian consistency condition
 \be \label{Wc} h_{\vec k}\dot h_{\vec k}^* - h_{\vec k}^*\dot h_{\vec k} = 2i \ . \ee 
 Substituting \eqref{phisolution2} into \eqref{motionscalar2} we get the equation
\be
\ddot{h}_{\vec{k}}+\left(m^2+\frac {\vec{k}^2}{a^2} -2  \frac{q A k_x}{a^2} +\frac{q^2 A^2}{a^2}+\left(6 \xi -\frac34\right)\frac{\dot a^2}{a^2}+\left(6 \xi -\frac32\right)\frac{\ddot a}{a}\right)h_{\vec{k}}=0  \ . \label{equhk2} \ee
\\

One can then write the vacuum expectation values for the stress-energy tensor as an integral over modes, producing an expression as
\be \label{integralT}\langle T^{\mu\nu} \rangle = \int d^3k T^{\mu \nu}(\vec k, t) \ . \ee 
In general, the above integral is ultraviolet divergent.  To renormalize (\ref{integralT}), 
 one should identify the subtraction terms that  can serve to remove the parts of $T^{\mu\nu}(\vec k, t)$  which would diverge when integrated over $\vec k$. A physically motivated procedure for doing this is to perform an adiabatic expansion of the mode functions $h_{\vec k} (t)$, in powers of $a(t)$, $\dot a(t)$, $\ddot a(t)$, etc and also in powers of $A(t)$, $\dot A(t)$, $\ddot A(t)$, etc, around the free  solutions. The mode expansion is then plugged in $T^{\mu\nu}(\vec k, t)$ to generate an adiabatic series.  The minimal number of terms in this series are subtracted from $T^{\mu\nu}(\vec k, t)$ to cancel out all UV divergences. This method was first developed in the context of a scalar field living in spatially homogeneous cosmological backgrounds \cite{parker-fulling, Bunch80, Anderson-Parker} and without any additional external field. 
 For more recent studies see \cite{rio1, LNT, RNT, Ghosh16, RFNT}.
  It was also applied, in the absence of gravity, to the pair-creation in a strong electric field scenario in \cite{Cooper1, Kluger91, Kluger92, Kluger93, anderson-mottola}. Here we will reexamine the method when both the gravitational and  the electric backgrounds are present. We will show that a consistent adiabatic order assignment is necessary for the method to agree with the combined conformal anomaly generated by the gravitational and the electromagnetic fields.

\subsection{Adiabatic expansion}

The adiabatic expansion for the scalar field modes is based on the usual WKB ansatz. For reasons to be explained later (see Sec. \ref{spinorqed}) we shall not assume {\it a priori} the WKB ansatz. Instead, we will assume a most general ansatz by expanding the modes as follows
\be h_{\vec{k}}=H_{\vec{k}}(t)e^{-i \int^{t}\Omega_{\vec k}(t')dt'} \ , \hspace{0.5cm}  \Omega_{\vec{k}}(t) = \omega_{\vec k} + \omega^{(1)} + \omega^{(2)} + \cdots \ , \hspace{0.5cm} H_{\vec{k}}(t)=\frac{1}{\sqrt{\omega_{\vec k}}}+H^{(1)}_{\vec{k}}(t)+\cdots \ee
where $\omega_{\vec k} = \sqrt{m^2 + \frac{\vec k^2}{a^2}}$ and $H_{\vec{k}}(t)$ and $\Omega_{\vec k}(t)$ are real functions.
One can substitute  the above ansatz into Eq. (\ref{equhk2}) and the Wronskian condition (\ref{Wc}). We then get the equations (we drop the $\vec{k}$ index for simplicity)
\bea
\ddot{H}-H \Omega^2 +\left[\omega^2-2\frac{ q A}{a}\frac{ k_x}{a} + \frac{q^2 A^2}{a^2}+\left(6 \xi -\frac34\right)\frac{\dot a^2}{a^2}+\left(6 \xi -\frac32\right)\frac{\ddot a}{a}\right]H&=&0 \nonumber  \\
\Omega H^2&=&1 \ . \label{system3q}
\eea
\\
We have to solve order by order to obtain the different terms of the expansion. As usual \cite{parker-toms}, we will consider $a(t)$ of adiabatic order zero, $\dot a(t)$ of adiabatic order one, etc.  However, to get an unique series expansion we have to assign also an adiabatic order to the vector potential function $A(t)$. {\it We will choose $A(t)$ to be of adiabatic order $1$}. 
This assignment of  adiabatic order 1 is consistent
with the scaling dimension of the field $A(t)$,  as it possesses the same dimensions as $\dot a$.
 The mass dimension of the scale
factor $a(t)$ is zero, while that of $\dot a(t)$, or the  field $A(t)$, is unity. [ We will reexamine this point in connection with the trace anomaly in subsection C.] 
Therefore, $\dot A (t)$ will be of adiabatic order $2$, $\ddot A(t)$ of order $3$ and so on. The adiabatic order $0$ coincides with the solution for $A(t)=0$:
\bea
\omega^{(0)}=\omega\\
H^{(0)}=\frac{1}{\sqrt{\omega}} \ , 
\eea
and hence
\be h^{(0)}_{\vec{k}}=\frac{1}{\sqrt{m^2 + \frac{\vec k^2}{a^2}}}e^{-i \int^{t} \sqrt{m^2 + \frac{\vec k^2}{a^2}} dt'} \ . \ee

\subsubsection{First adiabatic order}

By keeping only terms of first adiabatic order in \eqref{system3q}, the system of  two equations gives
\bea
-2 \omega \omega^{(1)}H^{(0)}-\frac{2 q A k_x}{a^2} H^{(0)}&=&0 \nonumber \ , \\*
2 \omega H^{(0)} H^{(1)}+\omega^{(1)}H^{(0)}H^{(0)} &=& 0 \ .
\eea
The solution is
\be H^{(1)} =\frac{k_x q A}{2 a^2 \omega^{5/2}} \ , \,\,\,\,\,\,\,\,\,\,\,\, \omega^{(1)} =-\frac{k_x q A}{a^2 \omega} \ .\ee

\subsubsection{Second adiabatic order}

In the same way, the second-order terms of (\ref{system3q}) give
\bea 
&&\ddot{H}^{(0)}-\left(2\omega \omega^{(2)}+\left(\omega^{(1)}\right)^2 +\frac{q^2 A^2}{a^2}+\left(6 \xi -\frac34\right)\frac{\dot a^2}{a^2}+\left(6 \xi -\frac32\right)\frac{\ddot a}{a}\right)H^{(0)}+\left(-2 \omega^{(1)}\omega-\frac{2 q A k_x}{a^2}\right)H^{(1)}= 0, \nonumber \\*
&&\omega^{(2)}\left(H^{(0)}\right)^2+2 \omega^{(1)}H^{(0)}H^{(1)}+\omega \left(H^{(1)}\right)^2+2 \omega H^{(2)}H^{(0)}=0 \ , 
\eea
which has as solutions
\bea \omega^{(2)}&=&\frac{3 \xi  \ddot{a}}{a \omega}-\frac{3 \ddot{a}}{4 a \omega}+\frac{3
   \xi  \dot{a}^2}{a^2 \omega}-\frac{3 \dot{a}^2}{8 a^2 \omega}
   -\frac{k_x^2 q^2 A^2}{2 a^4 \omega^3}+\frac{q^2 A^2}{2 a^2
  \omega}-\frac{\ddot{\omega}}{4\omega^2}+\frac{3
  \dot{\omega}^2}{8 \omega^3} \ , \nonumber \\*
H^{(2)} &=&\frac{5 k_x^2 q^2 A^2}{8 a^4 \omega^{9/2}}-\frac{q^2 A^2}{4 a^2
   \omega^{5/2}}-\frac{3 \xi  \ddot{a}}{2 a\omega^{5/2}}+\frac{3 \ddot{a}}{8 a\omega^{5/2}}-\frac{3 \xi  \dot{a}^2}{2 a^2 \omega^{5/2}}+\frac{3 \dot{a}^2}{16
   a^2 \omega^{5/2}}+\frac{\ddot{\omega}}{8 \omega^{7/2}}-\frac{3\dot{\omega}^2}{16 \omega^{9/2}} \ .\eea

\subsubsection{Third and fourth adiabatic order}

The same procedure can be repeated for all higher orders. The third- and fourth-order terms of the expansion are explicitly written in Appendix A.

\subsection{Conformal anomaly}

We check now the consistency of our adiabatic expansion. A nontrivial test for our proposal is  to reproduce the trace anomaly for the quantized charged scalar field for $\xi=1/6$ and $m=0$. 
To  evaluate the trace anomaly in the adiabatic regularization method, we have to start with a massive field and take the massless limit at the end of the calculation. Moreover,  for a massive charged field $ T^{\mu}_{\mu}= 2 m^2 \phi \phi^{\dagger}$. However, this formal identification does not imply that $ \langle T^{\mu}_{\mu} \rangle_{ren} = 2 m^2 \langle \phi \phi^{\dagger}\rangle_{ren}$. The divergences of the stress-energy tensor components have terms of fourth adiabatic order, while the divergences of $\langle \phi \phi^{\dagger}\rangle$ involve only terms till second adiabatic order. Therefore, in order to evaluate the trace anomaly by using the above formal expression, the adiabatic subtractions for $\langle  \phi \phi^{\dagger}\rangle$ should also include subtractions up to fourth adiabatic order. The same argument has been used to work out the trace anomaly of a real scalar field \cite{parker-toms}. 
Therefore, 
\be  \langle T^{\mu}_{\mu} \rangle_{ren}  = \lim_{m\to 0} 2m^2 ( \langle \phi\phi^{\dagger}\rangle_{ren}-\langle  \phi \phi^{\dagger}\rangle^{(4)}) \ . \ee
The fourth-order subtraction term, which produces a nonzero finite contribution when the mass vanishes, is codified in $\langle  \phi \phi^{\dagger}\rangle^{(4)}$. The piece $m^2 \langle  \phi \phi^{\dagger}\rangle_{ren}$ vanishes when $m^2 \to 0$. The remaining term produces the anomaly. We will now calculate this term. \\

Let us first consider the two-point function for our the complex scalar field
\bea
\langle\phi \phi^{\dagger}\rangle=\frac{1}{2(2\pi a)^3}\int |h_{\vec{k}}|^2 d^3k
\ , \eea
and, using the adiabatic expansion for the modes, one can evaluate the  corresponding $n$-order adiabatic terms
\bea
\langle \phi(t) \phi^{\dagger}(t)\rangle ^{(n)}=\frac{1}{2(2\pi a)^3}\int \left( |h_{\vec{k}}|^2\right)^{(n)} d^3k=
\frac{2 \pi}{2(2\pi a)^3}\int_{0}^{\infty}\int_{-\infty}^{\infty} |k_{\perp}| \left( |h_{\vec{k}}|^2\right)^{(n)} d k_x d k_{\perp} \ , 
\eea
where we have chosen a preferred direction $k_{\parallel}=k_x$.
 After some computation, the trace anomaly is finally given by
   \be \langle T^{\mu}_{\mu} \rangle_{ren} = \lim_{m^2 \to 0} - 2m^2\langle \phi \phi^{\dagger}\rangle^{(4)}= \frac{a^{(4)}}{240 \pi ^2 a}+\frac{\ddot a^2}{240 \pi ^2 a^2}+\frac{a^{(3)} \dot a}{80 \pi ^2
   a^2}-\frac{\dot a^2 \ddot a}{80 \pi ^2 a^3}-\frac{q^2 \dot A^2}{48 \pi ^2 a^2}\ . \ee   
   This last term is in full agreement with the well-known trace anomaly for a background electromagnetic field in Minkowski spacetime  \cite{arxiv}. The remaining terms reproduce  the trace anomaly of the gravitational background with FLRW metric $ds^2=dt^2-a^2(t)d\vec{x}^2$. The result is twice the value obtained for a real scalar field \cite{parker-toms}. In covariant form, we get
\bea
&& \left\langle T_{\mu}^{\mu}\right\rangle _{ren}= 	\frac{1}{1440 \pi^2}\left\{ \Box R- \left(R^{\mu\nu}R_{\mu \nu} - \frac13 R^2 \right) \right \} + \frac{q^2}{96 \pi^2}F_{\mu\nu}F^{\mu\nu} . \,\,\, \label{eq:Electroanomtrace}
\eea
The ability to reproduce the conformal anomaly is a nontrivial test for our renormalization scheme. 

\subsection{Discussion on the adiabatic order assignment}\label{discussion}

In most references in the literature \cite{Cooper1, Kluger91, Kluger92, Kluger93, anderson-mottola, KA, SSX} the vector potential is assumed of adiabatic order $0$, which differs from our previous assumption. This means that the adiabatic expansion proposed here and its physical consequences will be potentially different from those considered in the previous literature on this topic.  Assuming $A(t)$ of  adiabatic order zero, the leading adiabatic order of the modes would be
\be \tilde h^{(0)}_{\vec{k}}=\frac{1}{\sqrt{m^2+\frac{\vec{k}^2}{a^2}-\frac{2  q A k_x} {a^2}+	\frac{q^2 A^2}{a^2}} }e^{-i \int^{t} \sqrt{m^2+\frac{\vec{k}^2}{a^2}-\frac{2  q A k_x} {a^2}+	\frac{q^2 A^2}{a^2}} dt'} \ . \ee
A natural question now is to investigate whether this alternative adiabatic order  assignment,  is also able to reproduce the trace anomaly. The adiabatic order of the modes is given in Appendix B. 
When gravity is included, the result for the trace anomaly is problematic. Performing the adiabatic subtraction for the stress-energy tensor until the fourth adiabatic order, as usual for quantized fields interacting with gravity \cite{parker-fulling, Bunch80, Anderson-Parker, parker-toms, birrell-davies}, one gets for the trace anomaly $\left\langle T_{\mu}^{\mu}\right\rangle _{ren}= \lim_{m^2 \to 0} - 2m^2\langle \phi \phi^{\dagger}\rangle^{(4)}$. The result is 
 \bea \left\langle T_{\mu}^{\mu}\right\rangle _{ren}&=& \frac{1}{1440 \pi^2}\left\{ \Box R- \left(R^{\mu\nu}R_{\mu \nu} - \frac13 R^2 \right) \right \} \nonumber \\
 &-&\frac{q^2 \ddot a \dot A^2}{90 \pi ^2 m^2
   a^3}-\frac{7 q^2 \dot a^2 \dot A^2}{1440 \pi ^2
   m^2 a^4}-\frac{q^2 \dot a \dot A \ddot A}{720 \pi ^2 m^2 a^3}+\frac{q^2
   \ddot A^2}{160 \pi ^2 m^2 a^2}-\frac{7 q^4 \dot A^4}{1440 \pi ^2 m^4
   a^4}+\frac{q^2 A^{(3)} \dot A}{120 \pi ^2 m^2 a^2}
   \label{conf0}
\eea
The terms involving $A(t)$ and their derivatives do not reproduce the expected contribution to the trace anomaly $\frac{q^2}{96\pi^2}F_{\mu\nu}F^{\mu\nu} $. In fact these terms are infrared divergent as $m\to 0$.
To reproduce the electromagnetic piece of the trace anomaly (which is then of second adiabatic order), one is forced to remove gravity (i.e., take $a(t)=1$) and renormalize the stress-energy tensor until the second adiabatic order only. This way one would obtain
 \be \left\langle T_{\mu}^{\mu}\right\rangle _{ren}^{a=1}=-2\lim_{m^2 \to 0} m^2\langle \phi \phi^{\dagger}\rangle^{(2)} =\frac{q^2}{96\pi^2}F_{\mu\nu}F^{\mu\nu} \ . \ee
If one ignores the gravitational background and take $a(t)=1$, it is then  perfectly consistent to choose $A(t)$ of adiabatic order zero. However, if one includes gravity  the renormalization with the zeroth adiabatic order assignment for $A(t)$ seems to be not  consistent with  the trace anomaly.  This is an important novelty of this paper and we will go back to this point when consider  the trace anomaly of Dirac fermions.

\section{Spinor QED in two dimensions, adiabatic expansion, and renormalization} \label{spinorqed}

In two dimensions, the Dirac equation in  presence of an external homogeneous electric field and a background metric of the FLRW form $ds^2 = dt^2 -a^2(t)dx^2$ is
 \bea
 (i \underline{\gamma}^{\mu}\nabla_{\mu}-m)\psi=0\label{diraceq},
 \eea
 where $\nabla_{\mu} \equiv \partial_{\mu}-\Gamma_{\mu} -i q A_{\mu}$ and $\Gamma_\mu$ is the spin connection. $\underline{\gamma}^{\mu}(x)$ are the spacetime-dependent Dirac matrices satisfying the anticommutation relations $\{\underline{\gamma}^{\mu},\underline{\gamma}^{\nu}\}=2g^{\mu\nu}$. These  gamma matrices are related with the Minkowskian ones by $\underline{\gamma}^0 (t) = \gamma^0$ and $\underline{\gamma}^1 (t) = \gamma^1 / a(t)$, and the components of the spin connections are $\Gamma_0=0$ and $\Gamma_1=(\dot{a}/2) \gamma_0 \gamma_1$. Therefore,  $\underline{\gamma}^{\mu}\Gamma_{\mu}=-\frac{\dot{a}}{2a}\gamma_0$ and we fix a gauge for the potential as $A_{\mu}=(0,-A(t))$. The Dirac equation \eqref{diraceq} becomes 
  \bea
 \label{Dirac}\left(i \gamma^0 \partial_0 +\frac{i}{2}\frac{\dot{a}}{a}\gamma^0+\left(\frac{i}{a}\partial_1+\frac{q A_1}{a}\right)\gamma^1-m\right)\psi=0.
 \eea
From now on we will use  the Weyl representation (with $\gamma^5 \equiv \gamma^0\gamma^1$)
\bea
\gamma^0 = \scriptsize
\left( {\begin{array}{cc}
 0 & 1  \\
 1& 0  \\
 \end{array} } \right),\hspace{2cm} 
\gamma^1 = \scriptsize \left( {\begin{array}{cc}
 0 & 1  \\
 -1& 0  \\
 \end{array} } \right), \hspace{2cm} \gamma^5 = \scriptsize \left( {\begin{array}{cc}
 -1 & 0  \\
 0& 1 \\
 \end{array} } \right) 
 \ . \eea
 Expanding the field in  momentum modes $\psi(t, x)=\sum_{k}\psi_{k}(t)e^{ikx}$, Eq. (\ref{Dirac}) is converted into 
 \bea
 \left(\partial_0+\frac{\dot{a}}{2a}+\frac{i}{a}(k+q A)\gamma^5 +im\gamma^0\right)\psi_k=0 \ .
 \eea
 We can construct two independent spinor solutions
 \bea
 u_{k}(t, x)&=&\frac{e^{ikx}}{\sqrt{2\pi a}} \scriptsize \left( {\begin{array}{c}
 h^{I}_k(t)   \\
 -h^{II}_k (t) \\
 \end{array} }\right) \\
  v_{k}(t, x)&=&\frac{e^{-ikx}}{\sqrt{2\pi a}} \scriptsize \left( {\begin{array}{c}
 h^{II*}_{-k} (t)  \\
 h^{I*}_{-k}(t)  \\
 \end{array} } \right)
\ , \eea
where $ h^{I}_k(t)$ and $ h^{II}_k(t)$ are appropriate solutions of the equations 

\bea \label{system}
\dot{h}^{I}_k-\frac{i}{a} \left(k+qA\right)h^{I}_k-i m h^{II}_k=0\\
\dot{h}^{II}_k+\frac{i}{a}\left(k+qA\right)h^{II}_k-i m h^{I}_k=0 \ . 
\eea
\\

The normalization condition $|h^{I}|^2+|h^{II}|^2=1$  leads to the usual Dirac scalar products
\bea
(u_k,u_{k'})&&=\int dx a u_k^{\dagger} u_{k'}=\delta(k-k')\\
(v_k,v_{k'})&&=\int dx a v_k^{\dagger} v_{k'}=\delta(k-k')\\
(u_k,v_{k'})&&=\int dx a u_k^{\dagger} v_{k'}=0 \ . 
\eea
This condition guarantees the anticommutation relations for the creation and annihilation operators 
$B_k$ and $D_k$, defined by the expansion of the Dirac field operator in terms of the above spinors 
\bea
\psi(t, x)=\int dk \left[B_k u_k(t, x)+D^{\dagger}_k v_k(t, x)\right] \ . 
\eea
The usual equal-time anticommutation relation holds
\bea
\{\psi_{\alpha}(t, x),\psi^{\dagger}_{\beta}(t, y)\}=\delta(x-y)\delta_{\alpha \beta} \ . 
\eea

\subsection{Adiabatic expansion}

As explained above, the Dirac equation splits into a system of two coupled equation for $h^I_k$ and $h^{II}_k$.
Our aim here is to obtain a self-consistent adiabatic expansion for these two functions $h^I_k$ and $h^{II}_k$. Inspired by the adiabatic expansion for Dirac fermions in a four-dimensional FLRW spacetime \cite{LNT, RNT, Ghosh16, RFNT}, we propose the following ansatz (for simplicity in the notation we omit the index $k$ in the functions $F_k, G_k$ and $\Omega_k$)
\bea
h_k^{I}=\sqrt{\frac{\omega-k/a}{2 \omega}}F(t)e^{-i \int^t \Omega(t')dt'}\\
h_k^{II}=-\sqrt{\frac{\omega+k/a}{2 \omega}}G(t)e^{-i \int^t \Omega(t')dt'}
\ , \eea
where $\omega=\sqrt{m^2 + k^2/a^2}$.
We find 
\bea
(\omega-k/a)\left(\dot{F}-i\Omega F-\frac{i}{a}(k+q A)F \right)+\frac{k}{a}\frac{m^2}{2\omega^2}\frac{\dot{a}}{a}F+im^2 G=0  \\
(\omega+k/a)\left(\dot{G}-i\Omega G+\frac{i}{a}(k+q A)G \right)-\frac{k}{a}\frac{m^2}{2\omega^2}\frac{\dot{a}}{a}G+im^2 F=0 \label{fermion2d}
\eea
and the normalization condition
\bea 
\frac{\omega+k/a}{2 \omega} \left |G\right |^2+ \frac{\omega-k/a}{2 \omega}\left|F\right|^2=1.
\eea
As in previous sections, we have also assumed that $A(t)$ is of adiabatic order 1. The zeroth adiabatic order solution is the Minkowskian solution with vanishing electric field 
\bea
h^{I(0)}= \sqrt{\frac{\omega-k/a}{2 \omega}}e^{-i t \omega}\\
h^{II(0)}=-\sqrt{\frac{\omega+k/a}{2 \omega}}e^{-i t \omega}
\eea
\\
We expand the functions $F, G, \Omega$ adiabatically 
\bea
&&F=1+F^{(1)}+F^{(2)}+... \ , 
\\
&&G=1+G^{(1)}+G^{(2)}+... \ , 
\\
&&\Omega=\omega+\omega^{(1)}+\omega^{(2)}+... \ . 
\eea
We  recursively calculate order by order the higher-order adiabatic terms. We split the functions $F^{(n)}$ and $G^{(n)}$ into real and imaginary parts: $F^{(n)}=F_x^{(n)}+iF_y^{(n)}$, $G^{(n)}=G_x^{(n)}+i G_y^{(n)}$. The results up to second order areas follows.
\subsubsection{First adiabatic order}
By keeping only terms of first adiabatic order in (\ref{fermion2d}), the system of three equations gives
\bea
(\omega-k/a)\left(-i\omega F^{(1)}-i \omega^{(1)}-\frac{i}{a}k F^{(1)} -\frac{i}{a} q A\right)+\frac{k}{a}\frac{m^2}{2\omega^2}\frac{\dot{a}}{a}+im^2 G^{(1)}=0\\*
(\omega+k/a)\left(-i\omega G^{(1)}-i \omega^{(1)}+\frac{i}{a}k G^{(1)}+\frac{i}{a} q A \right)-\frac{k}{a}\frac{m^2}{2\omega^2}\frac{\dot{a}}{a}+im^2 F^{(1)}=0\\*
(\omega - k/a) (F^{(1)} + F^{(1) *} ) + (\omega + k/a ) (G^{(1)} + G^{(1) *} ) = 0 \ .
\eea
We now treat independently the real and imaginary parts. We obtain for the imaginary part
\bea
(\omega-k/a)\left(-\omega F_x^{(1)}- \omega^{(1)}-\frac{1}{a}k F_x^{(1)} -\frac{1}{a} q A\right)+m^2 G_x^{(1)}=0\\*
(\omega+k/a)\left(-\omega G_x^{(1)}- \omega^{(1)}+\frac{1}{a}k G_x^{(1)}+\frac{1}{a} q A \right)+m^2 F_x^{(1)}=0\\*
(\omega - k/a) (2 F_x^{(1)} ) + (\omega + k/a ) (2G_x^{(1)} ) = 0 \ .
\eea
The solution is
\be F^{(1)}_x = -\frac{q A (\omega+k/a)}{2a \omega^2} \ , \,\,\,\,\,\,\,\,\,\,\,\, G^{(1)}_x = \frac{q A ( \omega-k/a)}{2 a \omega ^2} \ , \,\,\,\,\,\,\,\,\,\,\,\, \omega^{(1)} = \frac{k q A}{a^2 \omega} \ .\ee
On the other hand, the real part of the system gives
\bea
(\omega-k/a)\left(\omega F_y^{(1)}+\frac{1}{a}k F_y^{(1)} \right)+\frac{k}{a}\frac{m^2}{2\omega^2}\frac{\dot{a}}{a}-m^2 G_y^{(1)}=0\\*
(\omega+k/a)\left(\omega G_y^{(1)}-\frac{1}{a}k G_y^{(1)} \right)-\frac{k}{a}\frac{m^2}{2\omega^2}\frac{\dot{a}}{a}-m^2 F_y^{(1)}=0
\ . \eea
These two equations are not independent. The obtained solution is
\be F^{(1)}_y = M(t) -\frac{k \dot a}{2a^2 \omega^2} \ , \,\,\,\,\,\,\,\,\,\,\,\, G^{(1)}_y = M(t) \ , \ee
where $M(t)$ is an arbitrary first-order adiabatic function. 
We have checked that physical expectation values of local observables are independent to any potential ambiguity in the choice for $M(t)$. For simplicity we choose 
\bea
M(t)=\frac{k \dot a}{4a^2\omega^2} \ , 
\eea
and then 
\be F^{(1)}_y = -\frac{k \dot a}{4 a^2\omega^2} =- G^{(1)}_y  \ . \ee

\subsubsection{Second adiabatic order}

In the same way, we solve the  second-order terms of (\ref{fermion2d}) and find

\bea F^{(2)}_x =&&\frac{A^2 k q^2}{2 a^3 \omega ^3}-\frac{5 A^2 m^2 q^2}{8 a^2 \omega ^4}+\frac{A^2 q^2}{2
   a^2 \omega ^2}-\frac{k \ddot a}{8 a^2 \omega ^3}+\frac{m^2 \ddot a}{8 a \omega
   ^4}-\frac{\ddot a}{8 a \omega ^2}+\frac{5 k m^2 \dot a^2}{16 a^3 \omega ^5}-\frac{k
   \dot a^2}{16 a^3 \omega ^3} \nonumber \\*&&-\frac{5 m^4 \dot a^2}{16 a^2 \omega ^6}+\frac{13 m^2
   \dot a^2}{32 a^2 \omega ^4}-\frac{3 \dot a^2}{32 a^2 \omega ^2} \ , \nonumber \\*
   G^{(2)}_x =&&\frac{A^2 k q^2}{2 a^3 \omega ^3}-\frac{5 A^2 m^2 q^2}{8 a^2 \omega ^4}-\frac{A^2 q^2}{2
   a^2 \omega ^2}+\frac{k \ddot a}{8 a^2 \omega ^3}+\frac{m^2 \ddot a}{8 a \omega
   ^4}-\frac{\ddot a}{8 a \omega ^2}-\frac{5 k m^2 \dot a^2}{16 a^3 \omega ^5}+\frac{k
   \dot a^2}{16 a^3 \omega ^3} \nonumber \\*&&-\frac{5 m^4 \dot a^2}{16 a^2 \omega ^6}+\frac{13 m^2
   \dot a^2}{32 a^2 \omega ^4}-\frac{3 \dot a^2}{32 a^2 \omega ^2} \ , \nonumber \\*
   \omega^{(2)} =&&-\frac{m^2 \ddot a}{4 a \omega ^3}+\frac{\ddot a}{4 a \omega}+\frac{5 m^4
   \dot a^2}{8 a^2 \omega ^5}-\frac{5 m^2 \dot a^2}{8 a^2 \omega a^3}+\frac{m^2
   q^2 A^2}{2 a^2 \omega ^3}\ ,\eea

\bea F^{(2)}_y =&&N(t)-\frac{5 m^2 q A \dot a}{4 a^2 \omega ^4}+\frac{3 q A \dot a}{4 a^2 \omega
   ^2}+\frac{q \dot A}{2 a \omega ^2} \ , \nonumber \\*
   G^{(2)}_y =&&N(t) \ .\eea
Again we have the same ambiguity for $N(t)$. We choose for simplicity:
\bea 
N(t)=\frac{5 m^2 q A \dot a}{8 a^2 \omega ^4}-\frac{3 q A \dot a}{8 a^2 \omega
   ^2}-\frac{q \dot A}{4 a \omega ^2} \ .
\eea
\subsection{Chiral anomaly}

To test the self-consistency of the above adiabatic expansion we are going to show how the chiral anomaly is obtained from it. We will consider the axial current
\bea
j_A^{\mu}= \bar{\psi}\gamma^{\mu}\gamma^5 \psi \ , 
\eea
which is conserved in the massless limit. To evaluate the expectation value  $\langle\nabla_{\mu}j_A^{\mu}\rangle$ we will reintroduce the mass and evaluate the 
right-hand-side of 
\bea
\langle \nabla_{\mu}j_A^{\mu}\rangle=2 i m \langle \bar{\psi}\gamma^5 \psi\rangle, \label{equaxial}
\eea
in the limit $m\to 0$. Since the formal expression for $\langle \nabla_{\mu}j_A^{\mu}\rangle$ has divergences till second adiabatic order we need to perform subtractions in $\langle \bar{\psi}\gamma^5 \psi\rangle$ up to second adiabatic order. Therefore,
\be 
\langle \nabla_{\mu}j_A^{\mu}\rangle_{ren}=-\lim_{m\to 0} 2 i m \langle \bar{\psi}\gamma^5 \psi\rangle^{(2)}  \ . \ee
By writing  $\langle \nabla_{\mu}j_A^{\mu}\rangle$  in terms of $\{h^{I},h^{II}\}$
\bea
 \langle \bar{\psi}\gamma^5 \psi \rangle=\frac{1}{2 \pi a} \int_{-\infty}^{\infty} dk (h^{II*}h^{I}-h^{I*}h^{II}) \ , 
\eea
and using our adiabatic series expansion, we arrive at
 \bea
 \langle \bar{\psi}\gamma^5 \psi \rangle^{(2)}=\frac{1}{2\pi a}\int^{\infty}_{-\infty} dk \frac{i m}{\omega}(F_y^{(2)}-G_y^{(2)}+F_y^{(1)}G_x^{(1)}-G_y^{(1)}F_x^{(1)})=\frac{i q \dot{A}}{2\pi a m} \ .
\eea
This result leads immediately to the axial anomaly in two dimensions
\be 
\langle \nabla_{\mu}j_A^{\mu}\rangle_{ren}=\frac{q \dot{A}}{a \pi}= -\frac{q}{2\pi}\epsilon^{\mu\nu}F_{\mu\nu}\ ,\ee 
where $\epsilon^{01}= |g|^{-1/2}= a^{-1} $.  This result reproduces exactly the chiral anomaly for spinor QED$_2$ \cite{Bertlmann, parker-toms}. For a massive field we obtain $
\langle \nabla_{\mu}j_A^{\mu}\rangle_{ren}= -\frac{q}{2\pi}\epsilon^{\mu\nu}F_{\mu\nu} + 2 i m \langle \bar{\psi}\gamma^5 \psi\rangle_{ren}$. The axial anomaly in two dimensions is the hallmark of the particle creation process caused by the electric field \cite{Blaer}. A manifestation of a similar phenomenon for photons in a gravitational scenario has been pointed out in \cite{ADNS}.

\subsection{Conformal anomaly}
It is very easy to see how the method accounts for  the trace anomaly. The trace of the energy momentum tensor can be written as:
\be  T^{\mu}_{\mu}  =m  \bar{\psi}\psi \ . \ee
After renormalization we have a residual contribution when the mass goes to zero
\be  \langle T^{\mu}_{\mu} \rangle_{ren} = \lim_{m\to 0} -m \langle \bar{\psi}\psi \rangle^{(2)} \label{tren} \ . \ee
By using the adiabatic expansion we can write:
\bea
\langle \bar{\psi}\psi \rangle^{(2)}=\frac{1}{2\pi a}\int_{-\infty}^{\infty} dk \left( h^{I*}h^{II}+h^{II*}h^{I}\right)^{(2)}=\frac{-1}{2\pi a}\int_{-\infty}^{\infty} dk \frac{m}{\omega}\left( F_x^{(2)}+G_x^{(2)}+F_x^{(1)}G_x^{(1)}+F_y^{(1)}G_y^{(1)}\right)
\eea
After integrating the corresponding adiabatic terms:
\bea
\langle \bar{\psi}\psi \rangle^{(2)}=\frac{\ddot{a}}{12\pi a m}.
\eea
By using \eqref{tren} we get
\bea \label{Ta}
\langle T^{\mu}_{\mu}\rangle_{ren}=-\frac{\ddot{a}}{12\pi a}
=-\frac{R}{24 \pi} \ , 
\eea
where in the last step we have used the expression of the two-dimensional scalar curvature in the terms of the expansion factor. The result agrees with the value of the trace anomaly for a Dirac spinor in two dimensions \cite{duff2}, which in turn coincides with the trace anomaly of a real scalar field \cite{birrell-davies, davies, Duff}.\\

We would like to stress that, if one assumes the zeroth adiabatic order for $A(t)$, the result for the trace anomaly is the following 
\bea
\langle T^{\mu}_{\mu}\rangle_{ren}=\lim_{m\to 0} [-\frac{\ddot{a}}{12\pi a}-\frac{q^2 \dot A^2}{6\pi m^2 a^2}] \ . 
\eea
We find again a very unpleasant result, like \eqref{conf0} for scalar fields in four-dimensions.  In the massless limit the above quantity is divergent and does not match  the expected result. By contrast, the adiabatic order one for $A$ gives the right result (\ref{Ta}) for the conformal anomaly.\\

\section{Renormalized current for a pulsed electric field in spinor QED$_2$}

\subsection{Vacuum choice for the Sauter pulse}
We consider now the  Sauter pulse in a two-dimensional Minkowski space. It is driven by the electric field $E= E_0 \cosh^{-2}(\omega_0 t)$, with potential $A(t)=-\frac{E_0}{\omega_0}(\tanh(\omega_0 t)+1)$. From the  Dirac equation we find the coupled equations:
\bea
\dot{h}^I-i P h^I- i m h^{II}&=&0 \\
\dot{h}^{II}+i P h^{II}-i m h^{I}&=&0 \ , 
\eea
where we have defined $P=k+q A(t)$. One can decouple these equations and obtain
\bea
\left(\partial_t^2-i \dot{P}+P^2+m^2\right)h^{I}=0\label{equmot1}\\
\left(\partial_t^2+ i \dot{P}+P^2+m^2\right)h^{II}=0 \ . 
\eea
The solution to the above equations, with the appropriated boundary conditions at $t=-\infty$,  are given in terms of hypergeometric functions  \cite{EWOT,cylinfunction,nikishov}
\bea
h^{I}&=& \sqrt{\frac{\omega-k}{2\omega}} F(a,b,c; \tau) \tau^{\alpha}(1-\tau)^{\beta} \\
h^{II}&=&-\sqrt{\frac{\omega+k}{2\omega}} F(a',b',c; \tau) \tau^{\alpha}(1-\tau)^{\beta} \ , 
\eea 
where $\tau= 1/2(1+\tanh(\omega_0 t))$, $a=-i q \frac{E_0}{\omega^2}+\alpha+\beta$, $b=1+i q \frac{E_0}{\omega^2}+\alpha+\beta$, $c=1+2\alpha$ and
\bea
\alpha=-\frac{i}{2\omega_0}\sqrt{k^2+m^2} \ , \ \ \ \ \ \ \  \beta=\frac{i}{2\omega_0}\sqrt{(k-2\frac{q E_0}{\omega_0})^2+m^2} \ .  
\eea
$a'$ and $b'$ are obtained from  $a, b$ by replacing $E_0\to-E_0$ and $k\to-k$. In the limit $t \to -\infty$ or $\tau \to 0$, both solutions have the asymptotic behavior corresponding to positive frequency modes
\bea
h^{I/II}\sim  \mp \sqrt{\frac{\omega \pm k}{2 \omega}}e^{- i \omega t} \ .
\eea

\subsection{Renormalization of the electric current}

\begin{figure}[htbp]
\begin{center}
\begin{tabular}{c}
\includegraphics[width=100mm]{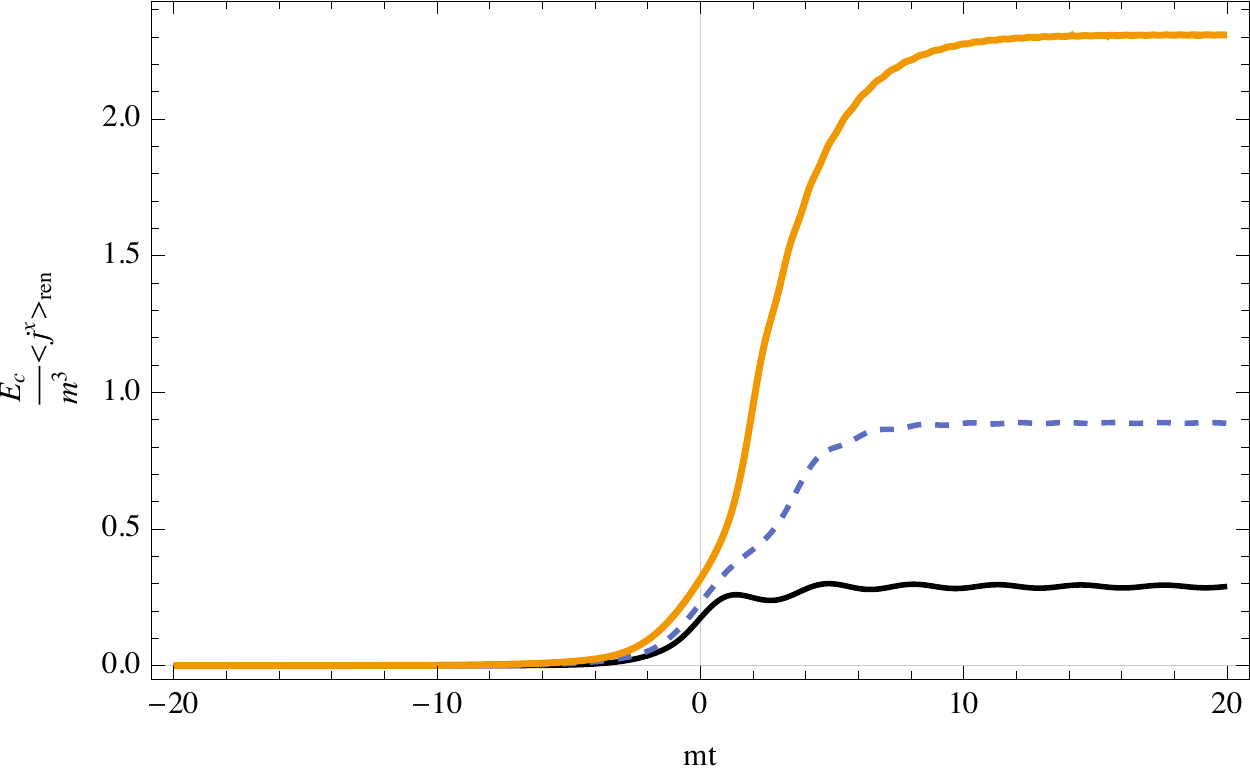}\\ 
\end{tabular}
\end{center}
\caption{\small{Renormalized electric current \eqref{jren} $\langle jx\rangle_{ren}$ for $E_0=2 E_c$ and $\omega_0=0.25 m$ (thick line), $\omega_0=0.35 m$ (dashed line) and $\omega_0=0.5m$ (black line) in natural units.}}
\label{jx15}
\end{figure}

The two-dimensional electric current defined as the source of the Maxwell equation: 
\bea
\partial_{\mu}F^{\mu\nu}= j^{\nu}
\eea

is  $j^{\mu}=-q\bar{\psi}\gamma^{\mu} \psi$. The formal expectation values $\langle j^\mu \rangle$, expressed in terms of $h^{I},h^{II}$, are given by  

\bea
\langle j^0 \rangle=-\frac{q}{2\pi} \int_{-\infty}^{\infty} dk (|h^{I}|^2+|h^{II}|^2) \\
\langle j^x \rangle=-\frac{q}{2\pi} \int_{-\infty}^{\infty} dk (|h^{I}|^2-|h^{II}|^2) \ .
\eea
For the $0$-component it is easy to see that after renormalization and taking into account the normalization condition, it vanishes $\langle j^0 \rangle_{ren}=0$.
In order to renormalize  the nonvanishing component of the electric current, the zeroth and first adiabatic orders need to be subtracted, i.e.,
\bea
\langle j^{x}\rangle_{ren}=\frac{q}{2\pi}\int_{-\infty}^{\infty} dk [|h^{II}|^2-|h^{I}|^2 -\left(|h^{II}|^2-|h^{I}|^2\right)^{(0)}-\left(|h^{II}|^2-|h^{I}|^2\right)^{(1)}].
\eea
After some calculations we find
\bea
\langle j^x \rangle_{ren}=\frac{q}{2\pi} \int_{-\infty}^{\infty} dk \left( |h^{II}|^2-|h^{I}|^2-\frac{k}{\omega}-\frac{q m^2} {\omega^3}A\right)\label{jren} \ .
\eea
For simplicity we assume $q>0$. The above integral is finite and one can estimate it easily by numerical integration (see Fig \ref{jx15}). We find a consistent  behavior for $\langle j^x \rangle_{ren}$. 
The induced current appears once the pulse starts and it depends on the values of the external parameters. $E_c \equiv m^2/q$ is the critical Schwinger value for pair production.  
Note  that for $\omega_0 \to 0$, i.e., the constant electric field configuration of the traditional Schwinger effect,  $\langle j^x \rangle_{ren} \to \infty$. We will interpret this behavior later on.  

\subsection{Adiabatic regime and particle creation}

Let us consider the adiabatic expansion for the renormalized electric current itself. In this case, after subtracting the zeroth and first adiabatic order, and assuming an adiabatic configuration for $A(t)$,  we find: 
\bea
\langle j^{x}\rangle_{ren}=\langle j^{x}\rangle^{(3)}+O^{(n>3)}=-\frac{7q^4 A^3}{30 m^2 \pi}-\frac{q^2 \ddot{A}}{6 m^2 \pi} +O^{(n>3)}\label{japrox3}
\eea
where $O^{(n>3)}$ are terms with adiabatic order higher than 3. We note first that, assuming that at early times the electric configuration is of the form $E(t) \geq 0$ and $\dot E(t) >0$, and taking into account that 
\be A(t) = -\int_{-\infty}^tdt' E(t')\ , \ee
the renormalized current  always takes positive values $\langle j^{x}\rangle_{ren} >0$ at early times, irrespective of the quantum state. \\

We can also compare the above  expansion with the exact result for the electric current for the Sauter pulse (see Fig. {\ref{comparision}). We observe that the configuration $A(t)$ defined by the Sauter pulse is only adiabatic at early times, and  ceases to be adiabatic around the instant $m t \approx -4$, time before reaching the critical electric field needed for pair production. 
The non-adiabatic period extends to all times in the (Schwinger) limit $\omega_0 \to 0$, and particles are created without end. Hence $\langle j^x \rangle_{ren}$ diverges, as note before. [Note that, in contrast, a configuration of the form $A(t)=-E_0(\tanh(\omega_0 t)+1)$, which corresponds to the electric field $E= E_0\omega_0$,  is adiabatic for $\omega_0 \to 0$ and the mean particle number $\langle N \rangle$ would be  zero, in agreement with the adiabatic theorem].

\begin{figure}[htbp]
\begin{center}
\begin{tabular}{c}
\includegraphics[width=100mm]{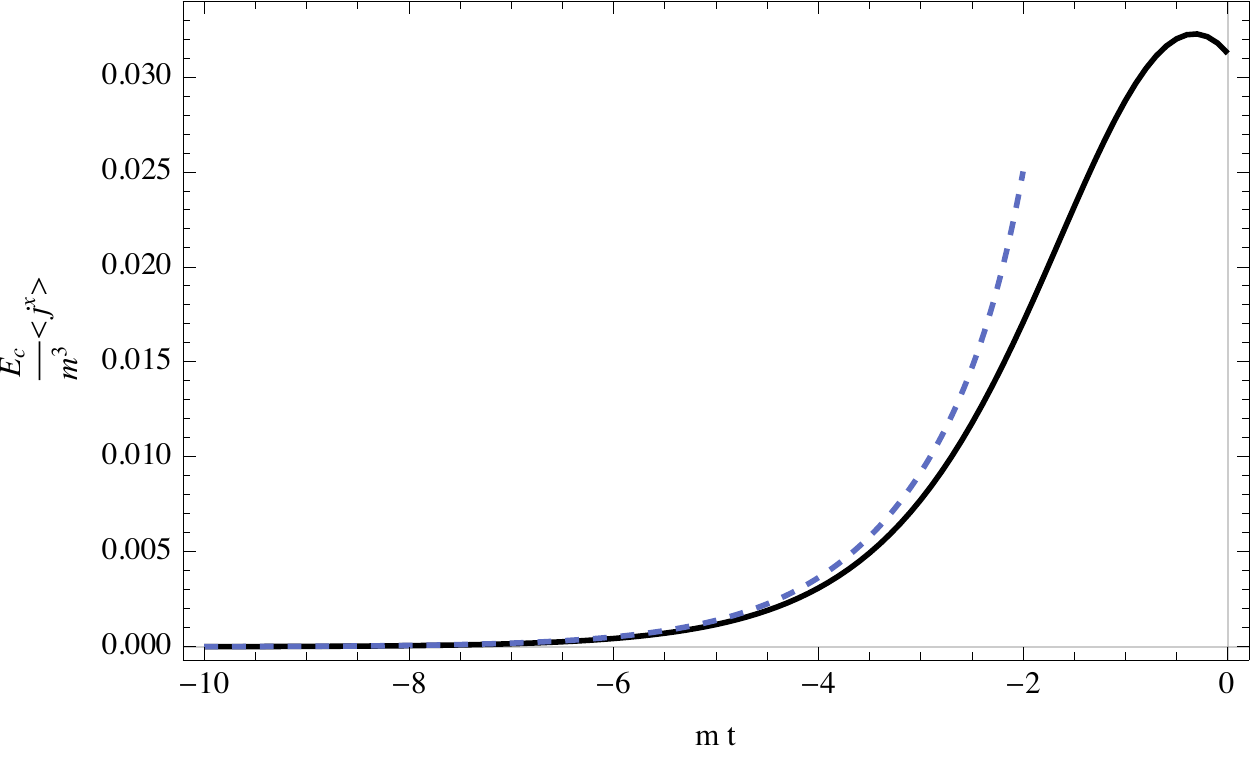}\\ 
\end{tabular}
\end{center}
\caption{\small{Renormalized electric current \eqref{jren} $\langle j^x\rangle_{ren}$ and third adiabatic order aproximation \eqref{japrox3} $\langle j_x\rangle^{(3)}$ for $E_0= E_c$ and $\omega_0=0.5 m$ in natural units.}}
\label{comparision}
\end{figure}

\subsection{Backreaction}

An important effect due to particle creation is the backreaction of the current production into the former electric field. Let us briefly illustrate this effect within our approach for renormalizing the electric current. In the case of the Sauter pulse the induced electric current behaves as a smooth function and we can mimic it with a  simple analytic function. Let us consider the pulse with $\omega_0=0.5 m$ and $E_0=4 E_c$. In this case we found that the electric current can be parametrize by a smooth function (see Fig. \ref{analytic}) such as:
\begin{figure}[htbp]
\begin{center}
\begin{tabular}{c}
\includegraphics[width=100mm]{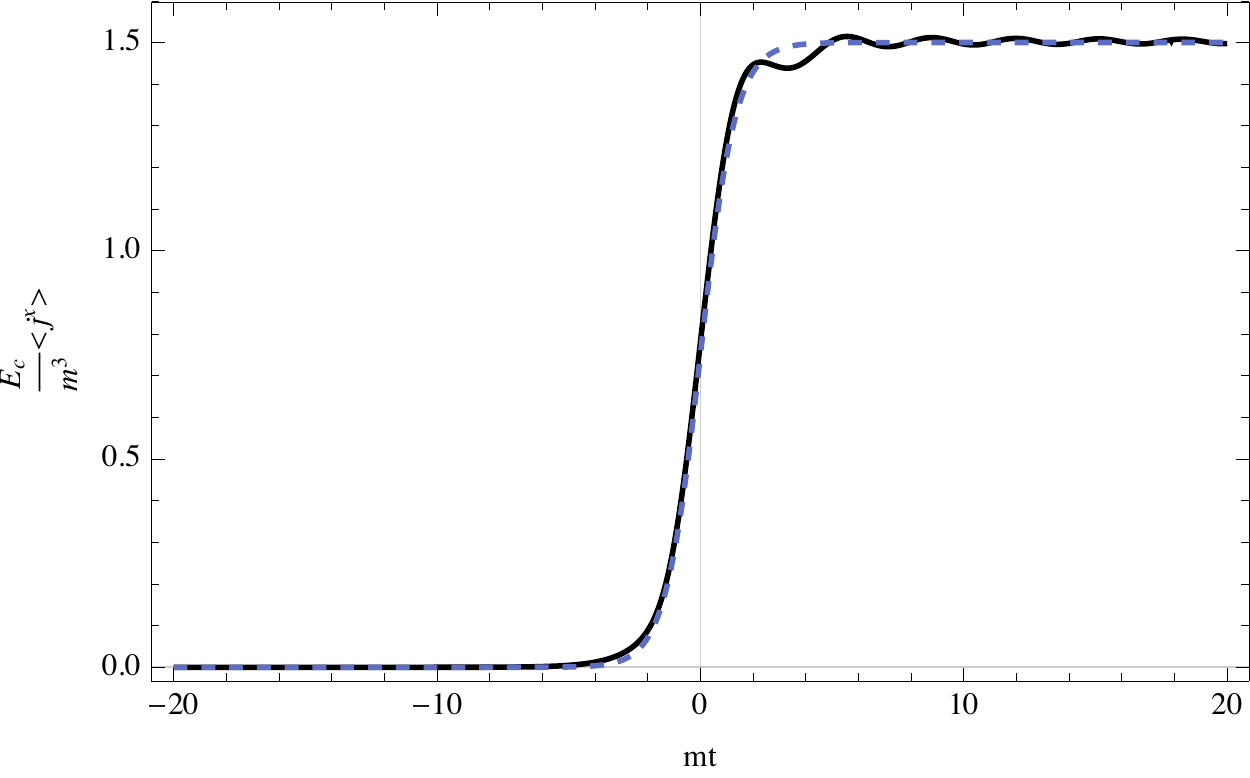}\\ 
\end{tabular}
\end{center}
\caption{\small{$\langle j^x\rangle_{ren}$ for $E_0=4E_c$ and $\omega_0=0.5 m$ in the case of numerical calculation (dashed line) and analytic function (black line).}}
\label{analytic}
\end{figure}

\bea
 \langle j^{x}\rangle_{ren}(t)=\frac{K m^3}{ E_c}(1+\tanh(K m t))  
\eea
where in this case $K=0.75$. The induced electric current also acts as source of the electromagnetic field according to the semiclassical  Maxwell equations:
\bea
\partial_{\mu}F^{\mu\nu}=\langle j^{\nu}\rangle_{ren}
\eea
In our case this translates into $\ddot{A}(t)=\langle j^{x}\rangle_{ren}(t)$. We can easily integrate this equation and find the backreaction contribution to the electric field $E(t)=-\dot{A}(t)$:
\bea
E_{br}(t)=\frac{m^2}{E_c} \left[-\log\left\{(2\cosh (K m t)\right\}-K m t\right]
\eea
where the integration constant is fixed in order to have vanishing potential/field at $t \to -\infty $.
\begin{figure}[htbp]
\begin{center}
\begin{tabular}{c}
\includegraphics[width=100mm]{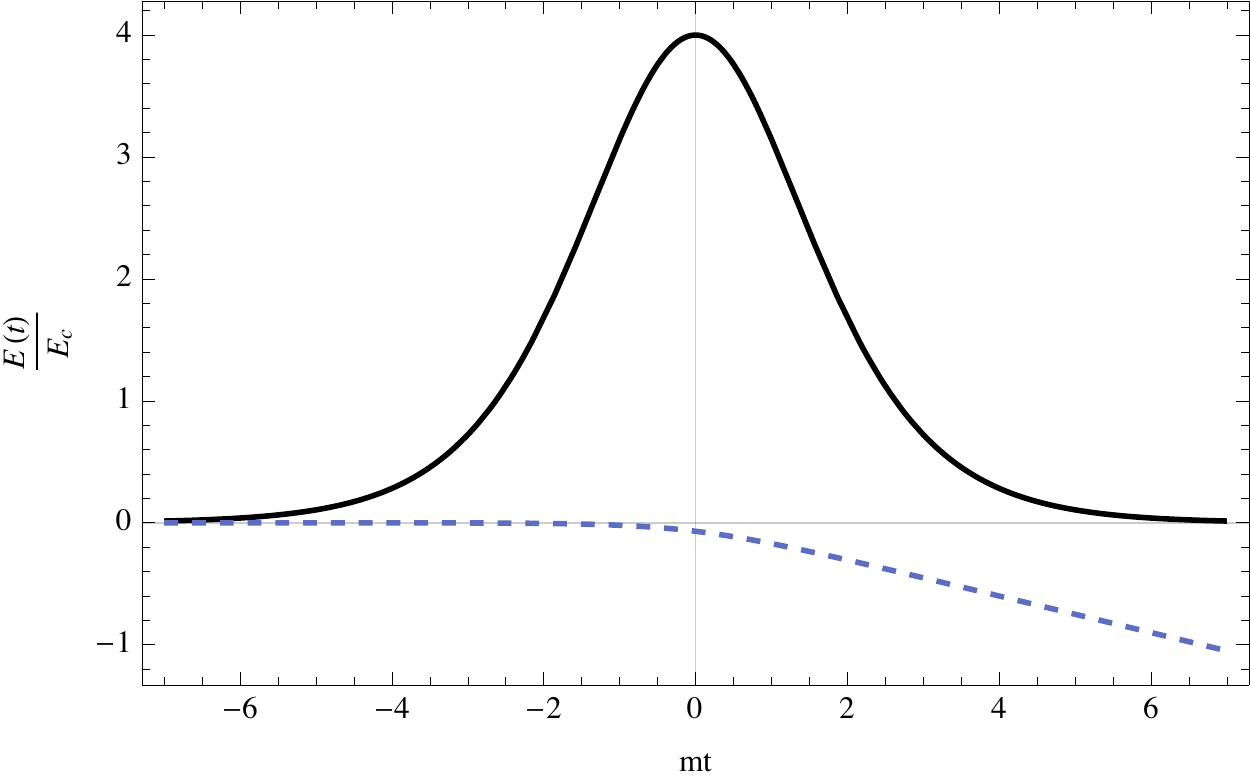}\\ 
\end{tabular}
\end{center}
\caption{\small{Original electric field $E(t)$(continuos line) and the electric field produced by particle creation (dashed line) in the case of $E_0=4 E_c$, $\omega_0=0. 5 m$
and a coupling of $q^2/m^2=0.1$}}
\label{backreaction}
\end{figure}

The electric field produced by backreaction to the particle creation becomes relevant once the pulse is almost over, when it reaches the critical electric field
$E_c$(see Fig. \ref{backreaction}).  It behaves now as a lineal function in time and goes in the opposite direction as the original electric field. The backreaction is significant around $mt\approx 3$.

\section{Summary}

When a quantum field is coupled to a classical, nonadiabatic time-dependent background, particles are produced. As a consequence new UV divergences in local observables emerge. In cosmological scenarios, adiabatic regularization provides a very efficient method to identify and remove the unwanted divergences. 
In the case of an electric field background with sufficient high intensity, pairs of particles and antiparticles are produced via the Schwinger effect. Equivalent to gravitational particle production, this generates divergences in quadratic observables such as the energy momentum tensor $\langle T_{\mu\nu} \rangle$  or the electric current $\langle j^{\mu}\rangle$ and a renormalization mechanism is needed in order to predict finite quantities. \\

In this paper we have improved  the adiabatic regularization method to include homogeneous electric fields. Our extension of the method has two folds. On the one hand, we have reexamined the method to deal with both electric and gravitational fields on an equal footing and for a quantized scalar field. In doing this we have fixed an inherent ambiguity of the method. The adiabatic order assignment of the vector potential has been traditionally assumed in the literature of zero order. Here we have argued that the correct adiabatic order assignment is one, instead of zero, at least if a gravitational field is present. This problem has been fixed by invoking the conformal anomaly. On the other hand, we have extended the adiabatic method to deal with fermions in two-dimensions. We have checked the consistency of our method by reproducing  both  the axial and  conformal anomalies. We have also shown that only the adiabatic order assignment one for $A$ yields the right expressions for the anomalies when gravity is present.  One of the main advantages of the adiabatic method in the capability to perform numerical computations. To briefly illustrate how to deal with the renormalized electric current predicted by our method we have analyzed numerically the electric current induced by a Sauter pulse.

\section*{Acknowledgments}

 We thank F. Barbero, S. Pla,  A. del Rio,  F. Torrenti and E. Villase\~nor for useful discussions. This work is partially funded  by the Grant. FIS2014-57387-C3-1-P, the COST action CA15117 (CANTATA), supported by COST (European Cooperation in Science and Technology), and the Severo Ochoa Program SEV-2014-0398.    A. F. is supported by the Severo Ochoa Ph.D. fellowship SEV-2014-0398-16-1 and the European Social Fund. 

\section*{APPENDIX A: ADIABATIC EXPANSION WITH $A(t)$ ASSUMED OF ADIABATIC ORDER ONE}\label{AppendixA}

In this appendix we provide the terms of the adiabatic expansion of the scalar field modes, up to fourth order, not included in the main text.
The third order terms are:
\bea &&\omega^{(3)} =-\frac{k_x^3 q^3 A^3}{2 a^6 \omega^5}+\frac{k_x q^3 A^3}{2 a^4 \omega
   ^3}+\frac{k_x q \ddot A}{4 a^2 \omega^3}-\frac{5 k_xq \dot A \dot \omega}{4 a^2
   \omega^4}-\frac{3 k_x q A \ddot \omega}{4 a^2 \omega^4}+\frac{19 k_x q A
   \dot \omega^2}{8 a^2 \omega^5}+\frac{3 k_x \xi  q A \dot a^2}{a^4 \omega^3} \nonumber \\&& +\frac{9 k_x q A \dot a^2}{8 a^4 \omega^3}+\frac{3 k_x \xi  q A
   \ddot a}{a^3 \omega^3}-\frac{5 k_x q A \ddot a}{4 a^3 \omega^3}-\frac{k_xq
   \dot a \dot A}{a^3 \omega^3}+\frac{5 k_x q A \ddot a \dot \omega}{2 a^3 \omega^4} \nonumber \ , \\
&&H^{(3)} =\frac{15 k_x^3 q^3 A^3}{16 a^6 \omega^{13/2}}-\frac{5  k_x q^3 A^3}{8 a^4 \omega
  ^{9/2}}-\frac{ k_x q \ddot A}{8 a^2 \omega^{9/2}}+\frac{5  k_x q \dot A \dot \omega
 }{8 a^2 \omega^{11/2}}+\frac{9  k_x q A \ddot \omega}{16 a^2 \omega
   ^{11/2}}-\frac{47  k_x q A \dot \omega^2}{32 a^2 \omega^{13/2}}-\frac{15 k_x
   \xi  q A \dot a^2}{4 a^4 \omega^{9/2}} \nonumber \\&&-\frac{9  k_x q  \dot a^2}{32 a^4 \omega
   ^{9/2}}-\frac{15  k_x \xi  q A \ddot a}{4 a^3 \omega^{9/2}}+\frac{19  k_x q
   A \ddot a}{16 a^3 \omega^{9/2}}+\frac{ k_x q \dot a \dot A}{2 a^3 \omega
   ^{9/2}}-\frac{5  k_x q A \dot a \dot \omega}{4 a^3 \omega^{11/2}} \ .  \eea

Finally, the fourth-order terms are:
\bea &&\omega^{(4)}=-\frac{q^4 A^4}{8 a^4 \omega^3}+\frac{3 k_x^2 q^4 A^4}{4 a^6 \omega^5}-\frac{5
    k_x^4 q^4 A^4}{8 a^8 \omega^7}-\frac{9 q^2 \dot a^2 A^2}{16 a^4 \omega
   ^3}-\frac{3 q^2 \xi  \dot a^2 A^2}{2 a^4 \omega^3}+\frac{103 k_x^2 q^2 \dot a^2
   A^2}{16 a^6 \omega^5}+\frac{9  k_x^2 q^2 \xi  \dot a^2 A^2}{2 a^6 \omega
  ^5}  \nonumber \\&&-\frac{19 q^2 \dot \omega^2 A^2}{16 a^2 \omega ^5}+\frac{145 k_x^2 q^2 \dot \omega
   ^2 A^2}{16 a^4 \omega^7}-\frac{5 q^2 \dot a \dot \omega A^2}{4 a^3 \Omega
   ^4}+\frac{25 k_x^2 q^2 \dot{a} \dot \omega A^2}{2 a^5 \omega^6}+\frac{5 q^2 \ddot a
   A^2}{8 a^3 \omega^3}-\frac{3 q^2 \xi  \ddot a A^2}{2 a^3 \omega^3}-\frac{21
   k_x^2 q^2 \ddot a A^2}{8 a^5 \omega^5} \nonumber \\&&+\frac{9 k_x^2 q^2 \xi  \ddot a A^2}{2
   a^5 \omega^5} +\frac{3 q^2 \ddot \omega A^2}{8 a^2 \omega^4}-\frac{15 k_x^2 q^2
   \ddot \omega  A^2}{8 a^4 \omega^6}+\frac{q^2 \dot a \dot A A}{a^3 \omega^3}-\frac{11
   k_x^2 q^2 \dot a \dot A A}{2 a^5 \omega^5} +\frac{5 q^2 \dot A \dot \omega A}{4 a^2
   \omega^4}-\frac{25 k_x^2 q^2 \dot A \dot \omega A}{4 a^4 \omega^6} \nonumber \\&&-\frac{q^2
   \ddot A A}{4 a^2 \omega^3} +\frac{3 k_x^2 q^2 \ddot A A}{4 a^4 \omega^5}-\frac{9
   \xi ^2 \dot a^4}{2 a^4 \omega^3}-\frac{27 \xi  \dot a^4}{8 a^4 \omega^3}+\frac{63
   \dot a^4}{128 a^4 \omega^3}-\frac{297 \dot \omega^4}{128 \omega ^7}-\frac{q^2 \dot A^2}{4 a^2
   \omega^3}+\frac{5 k_x^2 q^2 \dot A^2}{8 a^4 \omega^5}-\frac{57 \xi  \dot a^2 \dot \omega^2}{8 a^2 \omega^5} \nonumber \\ &&+\frac{57 \dot a^2 \dot \omega^2}{64 a^2 \omega^5}-\frac{9 \xi ^2
   \ddot a^2}{2 a^2 \omega^3}+\frac{3 \xi  \ddot a^2}{2 a^2 \omega^3}-\frac{9 \ddot a^2}{32 a^2
   \omega^3}-\frac{13 \ddot \omega^2}{32 \omega^5}-\frac{15 \xi  \dot a^3 \dot \omega}{2 a^3
   \omega^4}+\frac{15 \dot a^3 \dot \omega}{16 a^3 \omega^4} -\frac{9 \xi ^2 \dot a^2
   \ddot a}{a^3 \omega^3}+\frac{75 \xi  a^2 \ddot a}{8 a^3 \omega^3} \nonumber \\&&-\frac{27 \dot a^2
   \ddot a}{32 a^3 \omega^3}-\frac{57 \xi  \dot \omega^2 \ddot a}{8 a \omega^5}+\frac{57 \dot \omega^2 \ddot a}{32 a \omega^5}+\frac{15 \xi  \dot a \dot \omega \ddot a}{4 a^2 \omega^4}+\frac{9 \xi  \dot a^2 \ddot \omega }{4 a^2 \omega^4} -\frac{9 \dot a^2 \ddot \omega}{32 a^2
   \omega^4}+\frac{99 \dot \omega^2 \dot \omega}{32 \omega^6}+\frac{9 \xi  \ddot a \ddot \omega
   }{4 a \omega^4}-\frac{9 \ddot a \ddot \omega}{16 a \omega^4} \nonumber \\&&-\frac{3 \dot a
   a^{(3)}}{16 a^2 \omega^3}+\frac{15 \xi  \dot \omega \dddot{a}}{4 a \omega^4}-\frac{15
   \dot \omega \dddot{a}}{16 a \omega^4}-\frac{5 \dot \omega \dddot \omega}{8 \omega
   ^5}-\frac{3 \xi  \ddddot{a}}{4 a \omega^3} +\frac{3 a^{(4)}}{16 a \omega^3}+\frac{\ddddot \omega}{16 \omega^4}\ ,  \nonumber \\
    &&H^{(4)} =\frac{5 q^4 A^4}{32 a^4 \omega^{9/2}}-\frac{45 k_x^2 q^4 A^4}{32 a^6 \omega
  ^{13/2}}+\frac{195 k_x^4 q^4 A^4}{128 a^8 \omega^{17/2}}+\frac{9 q^2 \dot a^2
   A^2}{64 a^4 \omega^{9/2}}+\frac{15 q^2 \xi  \dot a^2 A^2}{8 a^4 \omega
   ^{9/2}}-\frac{457 k_x^2 q^2 \dot a^2 A^2}{128 a^6 \omega^{13/2}}-\frac{135
   k_x^2 q^2 \xi  \dot a^2 A^2}{16 a^6 \omega^{13/2}}  \nonumber \\&& +\frac{47 q^2 \dot \omega^2
   A^2}{64 a^2 \omega^{13/2}}-\frac{871 k_x^2 q^2 \dot \omega^2 A^2}{128 a^4 \omega
   ^{17/2}}+\frac{5 q^2 \dot a \dot \omega A^2}{8 a^3 \omega^{11/2}}-\frac{65 k_x^2
   q^2 \dot a \dot \omega A^2}{8 a^5 \omega^{15/2}}-\frac{19 q^2 \ddot a A^2}{32 a^3 \omega
   ^{9/2}}+\frac{15 q^2 \xi  \ddot a A^2}{8 a^3 \omega^{9/2}}+\frac{207 k_x^2 q^2
   \ddot a A^2}{64 a^5 \omega^{13/2}}  \nonumber \\&&-\frac{135 k_x^2 q^2 \xi  \ddot a A^2}{16 a^5
   \omega^{13/2}}-\frac{9 q^2 \ddot \omega  A^2}{32 a^2 \omega^{11/2}}+\frac{117
   k_x^2 q^2 \ddot \omega A^2}{64 a^4 \omega^{15/2}}-\frac{q^2 \dot a \dot A A}{2 a^3
   \omega^{9/2}}+\frac{7 k_x^2 q^2 \dot a \dot A A}{2 a^5 \omega^{13/2}}-\frac{5 q^2
   \dot A \dot \omega A}{8 a^2 \omega^{11/2}}+\frac{65 k_x^2 q^2 \dot A \dot \omega
   A}{16 a^4 \omega^{15/2}} \nonumber \\&&+\frac{q^2\ddot A A}{8 a^2 \omega^{9/2}}-\frac{9 k_x^2
   q^2 \ddot A \dot A}{16 a^4 \omega^{13/2}}+\frac{45 \xi ^2 \dot a^4}{8 a^4 \Omega
   ^{9/2}}+\frac{27 \xi  \dot a^4}{32 a^4 \omega^{9/2}}-\frac{99 \dot a^4}{512 a^4 \omega
   ^{9/2}}+\frac{621 \dot \omega^4}{512 \omega^{17/2}}+\frac{q^2 \dot A^2}{8 a^2 \omega
   ^{9/2}}-\frac{5 k_x^2 q^2 \dot A^2}{16 a^4 \omega^{13/2}} \nonumber \\&&+\frac{141 \xi  \dot a^2
   \dot \omega^2}{32 a^2 \omega^{13/2}}-\frac{141 \dot a^2 \dot \omega ^2}{256 a^2 \omega
   ^{13/2}}+\frac{45 \xi ^2 \ddot a^2}{8 a^2 \omega^{9/2}}-\frac{39 \xi  \ddot a^2}{16 a^2
   \omega^{9/2}}+\frac{45 \ddot a^2}{128 a^2 \omega^{9/2}}+\frac{29 \ddot \omega '^2}{128 \omega
   ^{13/2}}+\frac{15 \xi  \dot a^3 \dot \omega}{4 a^3 \omega^{11/2}}-\frac{15 \dot a^3 \omega
   }{32 a^3 \omega^{11/2}} \nonumber \\&&+\frac{45 \xi ^2 \dot a^2 \ddot a}{4 a^3 \omega^{9/2}}-\frac{231
   \xi  \dot a^2 \ddot a}{32 a^3 \omega^{9/2}}+\frac{81 \dot a^2 \ddot a}{128 a^3 \omega
   ^{9/2}}+\frac{141 \xi  \dot \omega^2 \ddot a}{32 a \omega^{13/2}}-\frac{141 \dot \omega ^2
   \ddot a}{128 a \omega^{13/2}}-\frac{15 \xi  \dot a \dot \omega \ddot a}{8 a^2 \omega
  ^{11/2}}-\frac{27 \xi  \dot a^2 \ddot \omega}{16 a^2 \omega^{11/2}}+\frac{27 \dot a^2 \ddot \omega
   }{128 a^2 \omega^{11/2}} -\frac{207 \dot \omega^2 \ddot \omega}{128 \omega
   ^{15/2}}\nonumber \\&&-\frac{27 \xi  \ddot a \ddot \omega}{16 a \omega^{11/2}}+\frac{27 \ddot a \ddot \omega
   }{64 a \omega^{11/2}} +\frac{3 \dot a \ddot{a}}{32 a^2 \omega^{9/2}}-\frac{15 \xi 
   \dot \omega \dddot a}{8 a \omega^{11/2}}+\frac{15 \dot \omega \dddot a}{32 a \omega
   ^{11/2}}+\frac{5 \dot \omega \dddot \omega}{16 \omega^{13/2}}+\frac{3 \xi  \ddddot{a}}{8
   a \omega^{9/2}}-\frac{3 \ddddot a}{32 a \omega^{9/2}}-\frac{\ddddot \omega}{32 \omega
   ^{11/2}} \ .\eea

\section*{APPENDIX B:ADIABATIC EXPANSION WITH $A(t)$ASSUMED OF ADIABATIC ORDER ZERO}
The adiabatic expansion of the modes is then 
\bea
\omega^{(0)}&=&\sqrt{\omega^2-\frac{2 q  k_x A}{a^2}+\frac{q^2 A^2}{a^2}}\equiv \Omega \hspace{1cm} H^{(0)}=\frac{1}{\omega^{(0)}}\\
\omega^{(1)}&=&0 \hspace{5cm} H^{(1)}=0 \\
\omega^{(3)}&=&0 \hspace{5cm} H^{(3)}=0 
\eea

\bea
&&H^{(2)}=\frac{-6 a\Omega^2 \ddot a-3 \Omega^2 \dot a^2-2 a^2 \Omega \ddot \Omega+3 a^2\Omega^2}{16 a^2\Omega^{9/2}}\ \hspace{1cm} \omega^{(2)}=-\frac{2 a \Omega ^2 \ddot a -\Omega^2 \dot a^2+2 a^2 \Omega \ddot \Omega-3 a^2 \dot \Omega^2}{8 a^2  \Omega^3}\\
&&H^{(4)}=\frac{3 \xi  \ddddot a}{8 a \Omega^{9/2}}-\frac{3 \ddddot a}{32 a \Omega
   ^{9/2}}-\frac{15 \xi  \dddot a \dot \Omega}{8 a \Omega^{11/2}}+\frac{15 \dddot a
   \dot \Omega }{32 a \Omega^{11/2}}+\frac{45 \xi ^2 \ddot a^2}{8 a^2 \Omega
   ^{9/2}}-\frac{27 \xi  \ddot a \ddot \Omega }{16 a \Omega^{11/2}}+\frac{141 \xi  \ddot a
   \dot \Omega ^2}{32 a \Omega^{13/2}}-\frac{39 \xi  \ddot a^2}{16 a^2 \Omega
   ^{9/2}}+\frac{27 \ddot a \ddot \Omega}{64 a \Omega ^{11/2}}\nonumber \\ &&-\frac{141 \ddot a \ddot \Omega^2}{128 a \Omega^{13/2}}+\frac{45 \ddot a^2}{128 a^2 \Omega^{9/2}}+\frac{45 \xi
   ^2 \dot a^4}{8 a^4 \Omega^{9/2}}-\frac{27 \xi  \dot a^2 \ddot \Omega}{16 a^2 \Omega
   ^{11/2}}+\frac{15 \xi  \dot a^3 \dot \Omega}{4 a^3 \Omega^{11/2}}+\frac{141 \xi  \dot a^2
   \dot \Omega^2}{32 a^2 \Omega^{13/2}}+\frac{27 \xi  \dot a^4}{32 a^4 \Omega
   ^{9/2}}+\frac{27 \dot a^2 \ddot \Omega}{128 a^2 \Omega^{11/2}}\nonumber \\ &&-\frac{15 \dot a^3 \dot \Omega}{32 a^3 \Omega^{11/2}}-\frac{141 \dot a^2 \dot \Omega^2}{256 a^2 \Omega
   ^{13/2}}-\frac{99 \dot a^4}{512 a^4 \Omega ^{9/2}}+\frac{3 a^{(3)} \dot a}{32 a^2
   \Omega^{9/2}}+\frac{45 \xi ^2 \dot a^2 \ddot a}{4 a^3 \Omega^{9/2}}-\frac{15 \xi  \dot a
   \ddot a \dot \Omega}{8 a^2 \Omega ^{11/2}}-\frac{231 \xi  \dot a^2 \ddot a}{32 a^3 \Omega
   ^{9/2}}+\frac{81 \dot a^2 \ddot a}{128 a^3 \Omega^{9/2}}-\frac{\ddddot \Omega}{32 \Omega
   ^{11/2}}\nonumber \\ &&+\frac{29 \ddot \Omega^2}{128 \Omega^{13/2}}+\frac{621 \dot \Omega^4}{512 \Omega
   ^{17/2}}+\frac{5 \dddot \Omega \dot \Omega}{16 \Omega^{13/2}}-\frac{207 \dot \Omega^2
   \ddot \Omega}{128 \Omega^{15/2}}\nonumber \\
&&\omega^{(4)}=-\frac{3 \xi  \ddddot a}{4 a \Omega ^3}+\frac{3 \ddddot a }{16
   a \Omega ^3}+\frac{15 \xi  \dddot a \dot \Omega }{4 a \Omega
   ^4}-\frac{15 \dddot a\dot \Omega }{16 a \Omega ^4}-\frac{9 \xi ^2
 \ddot a^2}{2 a^2 \Omega ^3}+\frac{3 \xi \ddot a^2}{2 a^2
   \Omega ^3} +\frac{9 \xi   \ddot a \ddot \Omega}{4 a \Omega ^4}-\frac{57 \xi 
   \ddot a \dot \Omega^2}{8 a \Omega ^5}-\frac{9 \ \ddot a^2}{32
   a^2 \Omega ^3}-\frac{9  \ddot a \ddot \Omega }{16 a \Omega ^4}+\frac{57
 \ddot a \dot \Omega ^2}{32 a \Omega ^5}\nonumber \\&&-\frac{27 \xi  \dot a^4}{8
   a^4 \Omega ^3}-\frac{9 \xi ^2 \dot a^4}{2 a^4 \Omega ^3}+\frac{63
   \dot a^4}{128 a^4 \Omega ^3}-\frac{15 \xi \dot a^3
   \dot \Omega }{2 a^3 \Omega ^4} +\frac{15  \dot a^3 \dot \Omega }{16 a^3 \Omega
   ^74}+\frac{9 \xi  \dot a^2 \ddot \Omega}{4 a^2 \Omega ^4}+\frac{15 \xi 
   \dot a \ddot a \dot \Omega}{4 a^2 \Omega ^4} -\frac{57 \xi  \
   \dot a^2 \dot \Omega^2}{8 a^2 \Omega ^5}-\frac{9  \dot a^2 \ddot \Omega}{32
   a^2 \Omega ^4}+\frac{57 \dot a^2 \dot \Omega ^2}{64 a^2 \Omega ^5}\nonumber \\ &&-\frac{9
   \xi ^2 \dot a^2 \ddot a}{a^3 \Omega ^3}+\frac{75 \xi   \dot a^2
   \ddot a}{8 a^3 \Omega ^3}-\frac{27  \dot a^2 \ddot a}{32 a^3 \Omega
   ^3}-\frac{3 \dddot a \dot a}{16 a^2 \Omega ^3}+\frac{\ddddot \Omega
 }{16 \Omega ^4}-\frac{13 \ddot \Omega ^2}{32 \Omega
   ^5}-\frac{297 \dot \Omega ^4}{128 \Omega ^7}-\frac{5 \dddot{\Omega}
   \dot{\Omega} }{8 \Omega ^5}+\frac{99  \dot \Omega^2  \ddot \Omega }{32 \Omega
   ^6}.\eea

\end{document}